\documentclass{article}
\usepackage{graphicx} 
\usepackage{subcaption}
\usepackage{listings}
\usepackage{xcolor}
\usepackage{url}
\usepackage{authblk}
\usepackage{bookmark}
\usepackage{rotating}
\usepackage{lscape}
\usepackage{float}
\PassOptionsToPackage{table}{xcolor}
\usepackage[table]{xcolor}
\usepackage{longtable}
\usepackage[title]{appendix}
\usepackage[font=small,labelfont=bf]{caption}
\usepackage{adjustbox}
\usepackage{booktabs}
\usepackage{changepage}
\usepackage{array}
\usepackage{multirow}
\usepackage{MnSymbol}
\usepackage{placeins}
\usepackage{amsmath}
\usepackage{gensymb}
\usepackage[margin=1.4in]{geometry}

\usepackage[numbers]{natbib}
\title{\centering Pressing Intensity: \\ An Intuitive Measure for Pressing in Soccer}
\author{} 
\date{}
\begin{document}

\maketitle
\vspace{-2em} 
\begin{center}
    {\Large Joris Bekkers} \\[1em]
    UnravelSports \\
    Breda, Netherlands \\
    \texttt{joris@unravelsports.com} \\[4em]
\end{center}

\begin{abstract}
Pressing is a fundamental defensive strategy in football, characterized by applying pressure on the ball owning team to regain possession. Despite its significance, existing metrics for measuring pressing often lack precision or comprehensive consideration of positional data, player movement, and speed. This research introduces an innovative framework for quantifying pressing intensity, leveraging advancements in positional tracking data and components from Spearman’s Pitch Control model. Our method integrates player velocities, movement directions, and reaction times to compute the time required for a defender to intercept an attacker or the ball. This time-to-intercept measure is then transformed into probabilistic values using a logistic function, enabling dynamic and intuitive analysis of pressing situations at the individual frame level. The model captures how every player’s movement influences pressure on the field, offering actionable insights for coaches, analysts, and decision-makers. By providing a robust and interpretable metric, our approach facilitates the identification of pressing strategies, advanced situational analyses, and the derivation of derived metrics, advancing the analytical capabilities for modern football.
\newline
\newline
\noindent
The full implementation of the the Pressing Intensity algorithm is available as open-source code through the \textbf{\textit{unravelsports}} \cite{unravelsports} Python package, introduced by Bekkers \& Sahasrabudhe (2024) \cite{bekkers2024graph}.

\end{abstract}

\section{Introduction}
\label{intro}
Pressing is defined by renowned football analysis blog \textit{Spielverlagerung} as creating tension with the intention of getting the ball back~\cite{basic17}. They state pressing is the application of pressure with a specific intent. Every movement on the pitch creates some sort of pressure or tension somewhere on the field. 
When pressing, a team is actively trying to win the ball back through pressuring the opponent and by moving out of or within its formation. 

Over the years many different ways to measure pressing have been attempted. \textit{Passes Allowed Per Defensive Action} (PPDA)~\cite{trainor14} creates a proxy for pressure using on-ball event data, where there is an inherit lack of positional information for non-on-ball players. This is done by creating a ratio of passes made by the attacking team divided by the number of defensive actions (tackles, fouls, interceptions and challenges) the defending team makes in the build-up zone. Another attempt to measure pressing is a rule-based approach introduced by data provider \textit{StatsBomb}. They include a pressure event for every player within a 4 to 5 yards radius of the ball carrier~\cite{morgan18}. ~\cite{robberechts19} introduces a novel metric that quantifies the effectiveness of pressing in different game scenarios as a trade-off between the benefits of recovering the ball versus the cost of leaving the defensive structure using on-ball event data and a model to compute the probability of a ball recovery by the pressing team and that of the attacking team scoring on the pressing team.
~\cite{merckx21} propose to automate the analysis of a soccer team’s defensive pressing strategy from a combination of positional and event data by detecting pressing situations using a set of expert-defined rules. These rules are focused around proximity, positioning of defenders and movement towards the ball.
~\cite{andrienko17} create a formula for numeric expression of the level of pressure that takes into account relative position of a defense player with respect to the target of the pressure (i.e., the ball or an opponent player) and the defended goal. Unfortunately, leaving out one of the more important aspects of pressure, namely speed. ~\cite{bauer21} automatically identify pressing strategies from 20,000 manually labeled transition situations in order to derive metrics that support coaches with the analysis of these transition situations.
Finally, although not directly related to pressing ~\cite{fernandez18} propose the concept of \textit{Space Occupation Gain} (SOG) as the relative amount of quality of owned space (using a combination of \textit{Pitch Control} and a model to value the controlled space) earned during a time window.

This research introduces an intuitive formulaic expression for pressing, derived from~\cite{spearman17} \textit{Pitch Control} model, that aims to measure pressing such that it can easily be understood, be used by coaches, assistants and (data) analysts to identify and analyze pressing situations, compute advanced derived metrics (e.g. moving out from under pressure after a receival), and analyze specific in-game situations related to pressing. This approach utilizes positional tracking data, player direction of movement, and their speed to determine the amount of pressure from every defensive player to every attacking player at the individual frame level. And as such, allows us to describe every movement on the pitch and how it creates some sort of pressure or tension somewhere on the field.

\section{Pressing Intensity}
Our approach to measure this \textit{Pressing Intensity} is constructed with components from Spearman's~\cite{spearman17} \textit{Pitch Control} model, Shaw's simplified implementation of this model~\cite{shaw20} and Pleuler's modification of Shaw's model~\cite{pleuler20}. An important aspect within this model is the \textit{expected time to intercept}, described by Spearman as the time it would take player $j$ to reach location $\vec{r}$ from their starting location $\vec{r}_{j}(t)$ with a starting velocity $\vec{v}_{j}(t)$ assuming they are able to accelerate with some constant acceleration $a$ to a maximum speed $v$. Shaw simplified it so that the \textit{time to intercept} - or the time it takes for a player to get to a location on the pitch - assumes the player continues to move at starting velocity $\vec{v}_{j}(t)$ for reaction time ($\tau_{r}$) seconds and then runs at full speed ($v_{max}$) to the target position ($\vec{r}_{target}$). Pleuler incorporates a parameter ($\tau_{\theta}$) that penalizes velocities facing away from the target location.





\noindent
Because \textit{pressing} is concerned with applying pressure to opponents and / or the ball, we update this formula to represent the \textit{time to intercept an opponent or ball} instead of the time to takes for a player to get to a location on the pitch.

\subsection{Time to Intercept a Moving Object}
The \textit{time to intercept an opponent or ball} ($T_{i,j}$) estimates the time required for a defending player $i$ (the player applying pressure) to intercept an attacking target $j$ based on their respective locations ($\vec{r}_{i}(t)$ and $\vec{r}_{j}(t)$) and velocities ($\vec{v}_{i}(t)$ and $\vec{v}_{j}(t)$) at time $t$, a similar reaction time parameter, a similar maximum running velocity, an updated penalty parameter ($\tau_{\beta}$) and a small positive value (e.g. $\epsilon=10^{-5}$) as described in Formula~\ref{eq:tti}). 

\begin{equation}
T_{i,j}(t) = \tau_{r} + \tau_{i,j}(t) + \tau_{\beta}(t)
\label{eq:tti}
\end{equation}

\begin{equation}
\tau_{i,j}(t) = \frac{\|\vec{d}\|}{v_{\text{max}}}
\quad \text{where} \quad
\begin{aligned}
\vec{d}_j &= \vec{r}_{j}(t) + \vec{v}_{j}(t) \\
\vec{d}_i &= \vec{r}_{i}(t) +  \vec{v}_{i}(t) \cdot \tau_r\\
\vec{d} &= \vec{d}_j - \vec{d_i}
\end{aligned}
\label{eq:tti-obj}
\end{equation}

\begin{equation}
\tau_{\beta}(t) = \frac{\|\vec{u}\| \cdot \beta}{\pi}
\quad\text{where } \quad 
\begin{aligned}
\vec{u} &= (\vec{r}_{i}(t) + \vec{v}_{i}(t)) - \vec{r}_{i}(t) \\
\vec{v} &= \vec{d}_j - \vec{r}_{i}(t) \\
\beta &= \arccos\left(\frac{\vec{u} \cdot \vec{v}}{\|\vec{u}\| \cdot \|\vec{v}\| + \varepsilon}\right)
\end{aligned}
\label{eq:tti-beta}
\end{equation}

\subsection{Probability to Intercept a Moving Object}
Now that we can compute the time to intercept from every defending player to each attacking player (and the ball) we convert this into probabilities by passing it through the logistic function, as proposed by Spearman~\cite{spearman17}. This now gives us the chance that a defending player will reach an attacking player or the ball in some amount of time ($T$) given both their locations and velocities as shown in Formula~\ref{eq:tti-sigmoid}. Here, $\sigma$ is set to 0.45 and $T$ is set to 1.5 seconds.

\begin{equation}
p_{i,j}(T_{i,j}(t), T  |  \sigma) = \left[1 + \exp\left(-\frac{\pi}{\sqrt{3} \cdot \sigma} \cdot (T - T_{\text{i,j}}(t))\right)\right]^{-1}
\label{eq:tti-sigmoid}
\end{equation}

\subsection{Total Pressure on an Object}
Now we can describe the total pressure on object ($P_{j}$) with Formula~\ref{eq:tti-sum}. This can be read as one minus the chance that none of the players $i$ on defending team can get to object $j$.
\begin{equation}
P_{j} = 1 - \prod_{i} \left( 1 - p_{i,j} \right)
\label{eq:tti-sum}
\end{equation}
With this formulation we make the naive assumption that all these probabilities are independent.

\subsection{A Visual Representation}
\begin{figure}[!ht]
    \centering
    \includegraphics[width=0.9\linewidth]{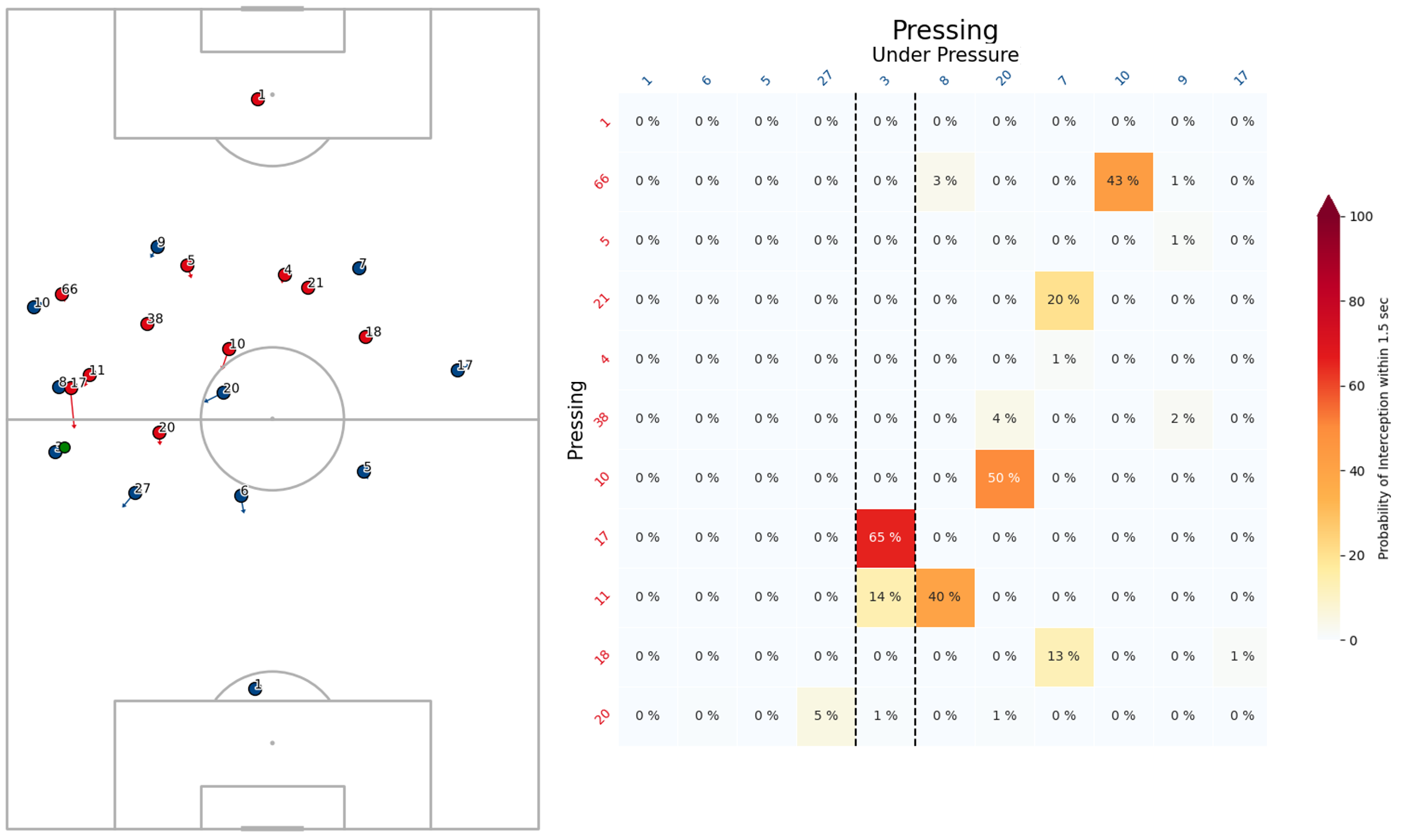}
    \caption{A visual representation of the pressure applied by every defender to every attacker}
    \label{fig:p}
\end{figure}

Figure~\ref{fig:p} shows a visual representation of the pressure applied by every defender to every attacker. In this example, and as an improvement to the proposed solution above, the pressure on the ball carrier (\#3, denoted in the column with the dashed boarder) is calculated as the maximum pressure applied by player $i$ on player \#3 and the ball. 

\section{Model Improvements}
We need to make some improvements to be able to practically use this model. One of the main issues with this implementation is that there is no distinguishing factor between an attacking player moving towards a defending player, or (what we are actually looking for) a defending player moving towards an attacking player. 

\subsection{Active Pressing}
\begin{figure}[!ht]
    \centering
    \includegraphics[width=0.9\linewidth]{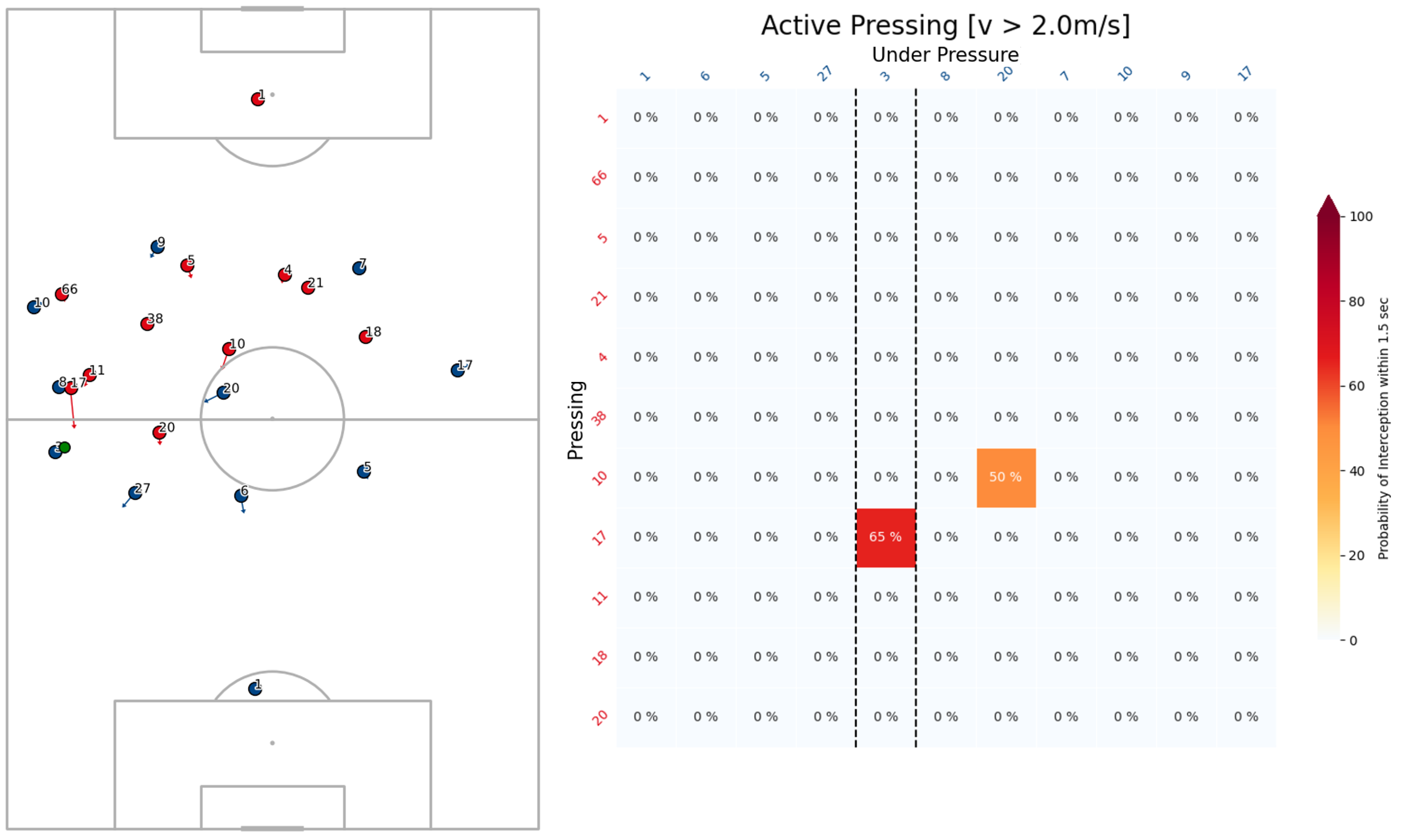}
    \caption{Filtered Pressing Intensity with a speed threshold set at 2 m/s}
    \label{fig:p1}
\end{figure}

To account for this we introduce an \textit{Active Pressing} speed threshold such that any chance of intercepting an attacking player is set to zero when a defending players’ speed is below this threshold. Figure~\ref{fig:p1} shows the unfiltered pressing intensity for a snapshot of tracking data, and Figure 4 shows the same situation with the noise filtered out through this speed threshold.
\subsection{Pitch Boundaries}
\begin{figure}[!ht]
    \centering
    \includegraphics[width=0.9\linewidth]{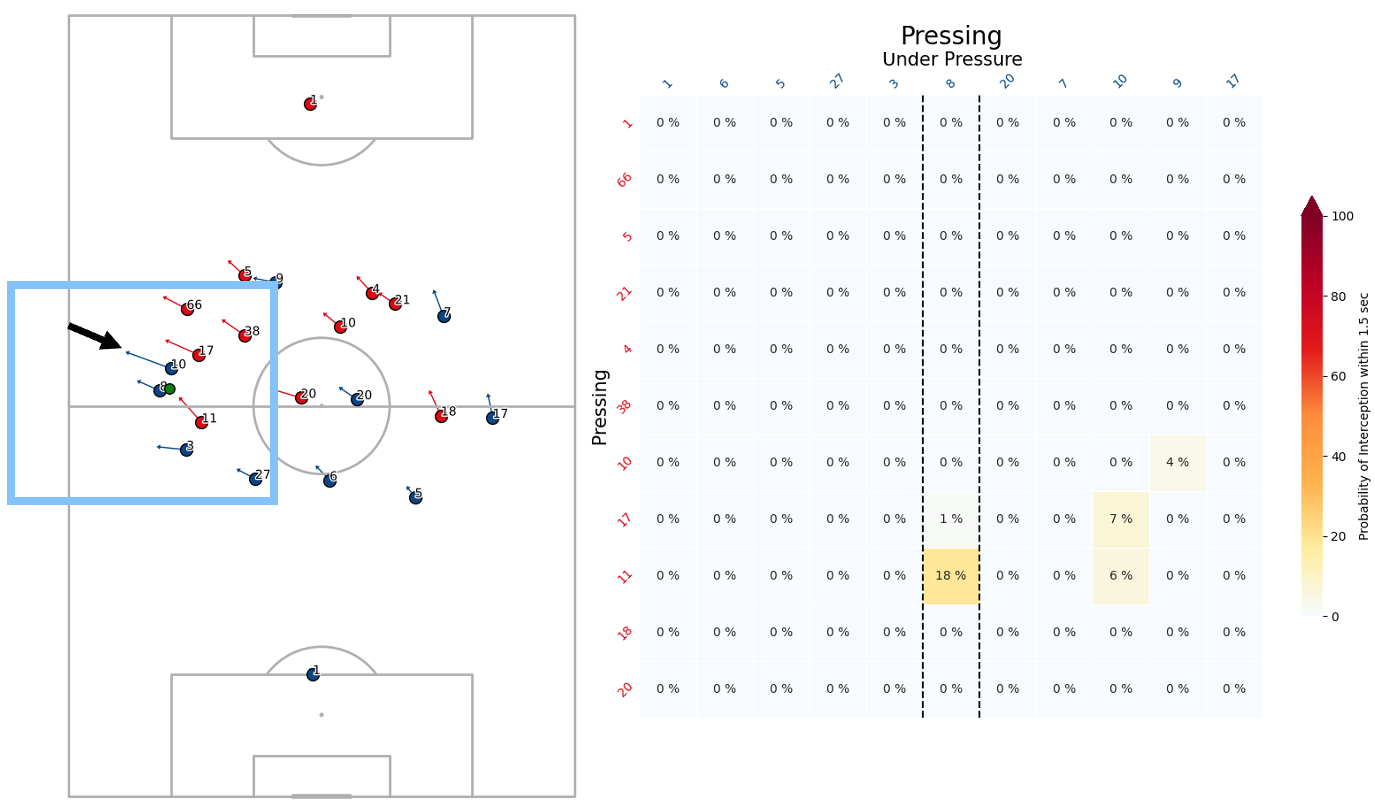}
    \caption{\centering No pressure in the system - even though we would expect it - because players are being forced toward the sideline, but are impossible to catch up to.}
    \label{fig:pb}
\end{figure}

An issue with this approach is shown in Figure~\ref{fig:pb}. The model assumes an infinite playing surface. Consequently, situations where a player is being pressed towards the sideline but cannot be caught within 1.5 seconds are treated as pressure free.

This limitation could be addressed by introducing a Pressing Intensity along the sidelines, acting opposite and equal to the force exerted by the attacking player(s) moving toward it.

\section{Future Work}
This Pressing Intensity model can be extended for other use cases, two of these are outlined below.
\subsection{Pressing / Closing Down Pass Lanes}
\begin{figure}[!ht]
    \centering
    \includegraphics[width=0.4\linewidth]{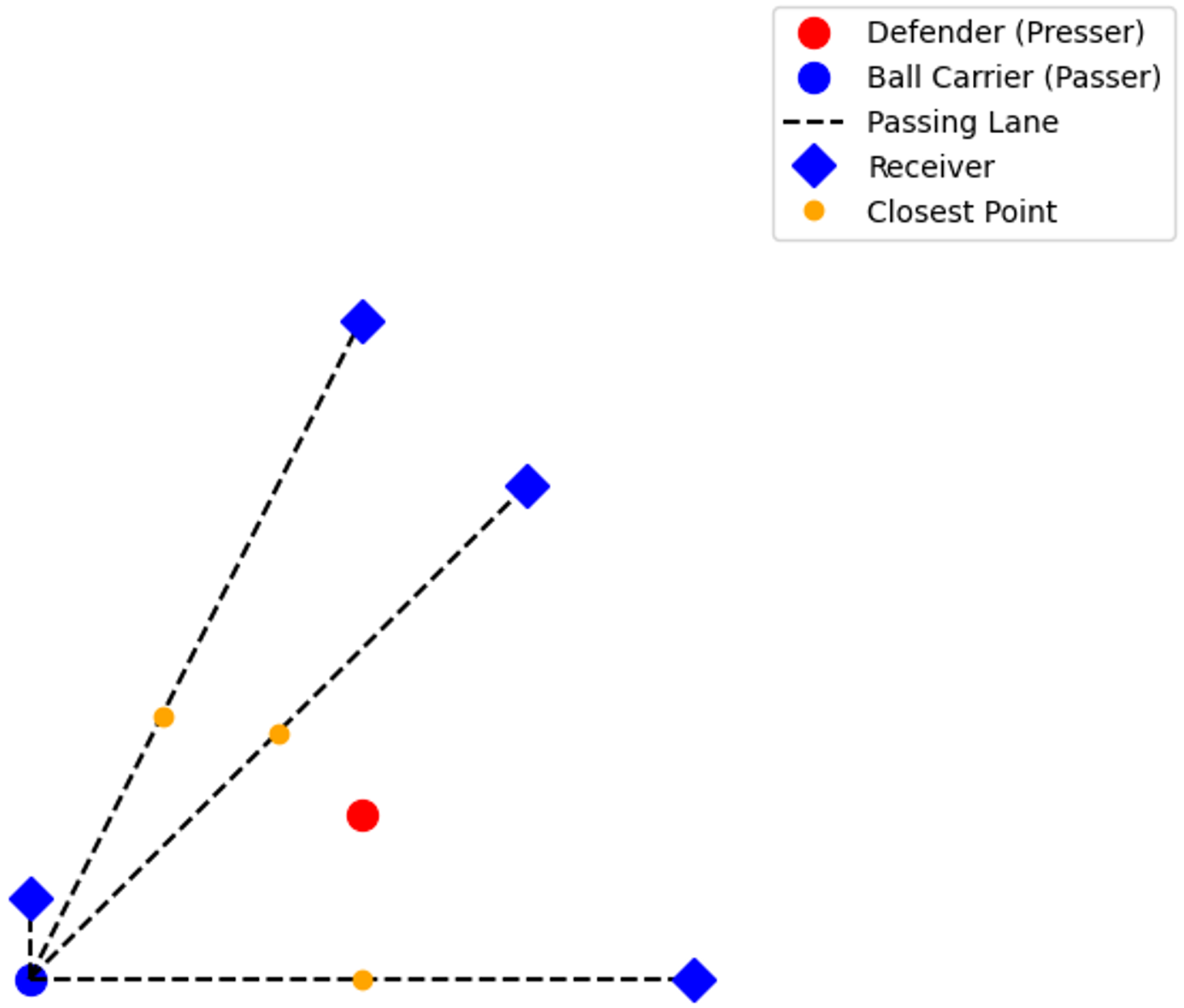}
    \caption{\centering A simple sketch of the closest point (in orange) for the pressing player (in red) to block passing options of the Blue team.}
    \label{fig:ppd}
\end{figure}

By replacing the locations and velocities of the attacking players by the closest point (and their respective velocities) of each defensive player towards each pass lane (see Figure~\ref{fig:ppd}) we can measure Pressing Intensity in relation to closing down passing options, instead of closing down the attacking players directly.

\begin{figure}[!ht]
    \centering
    \includegraphics[width=0.9\linewidth]{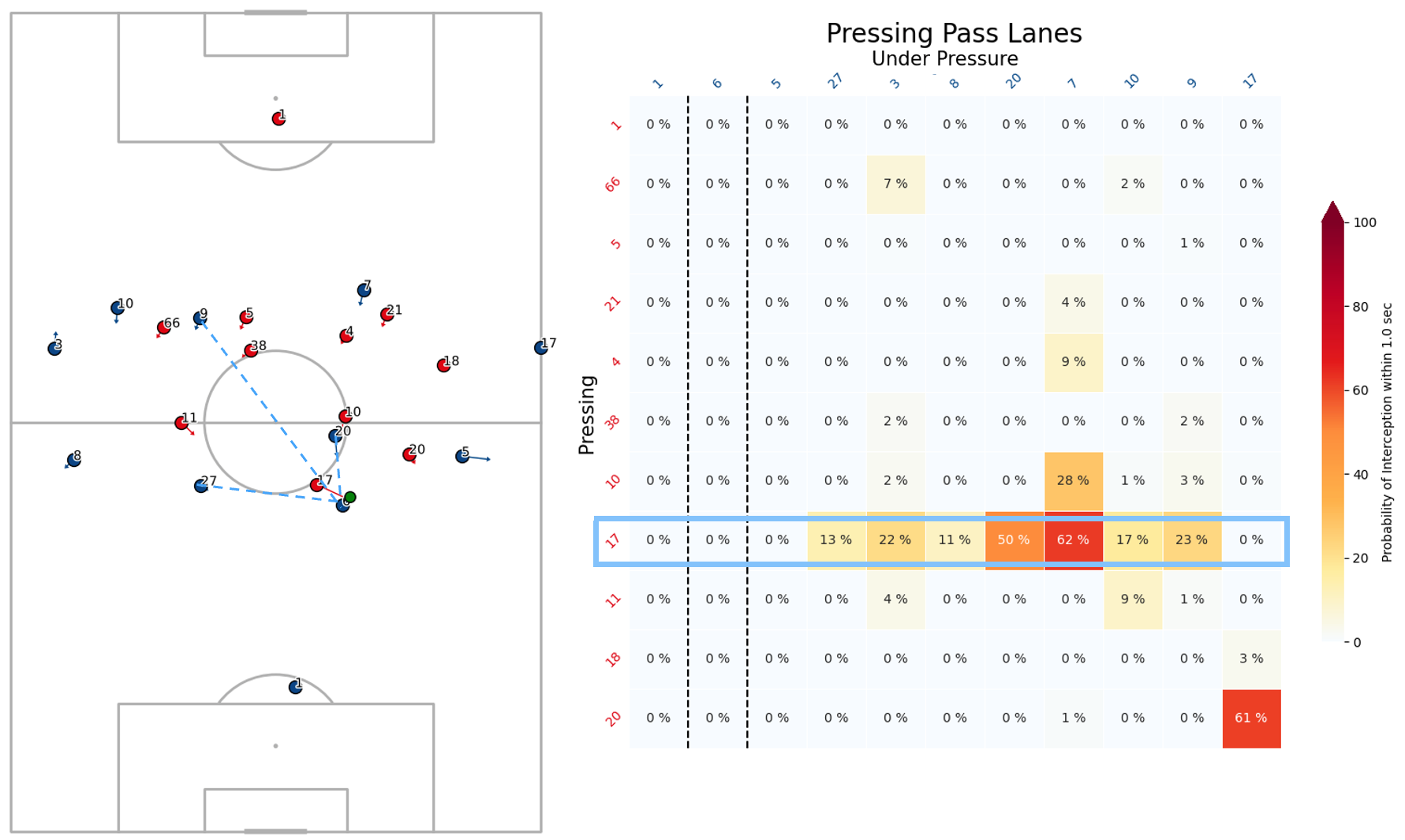}
    \caption{\centering An example of pressing passing lanes, instead of players. In this example $T$=1 second.}
    \label{fig:pl}
\end{figure}

An example of this is shown in Figure~\ref{fig:pl}. This implementation can be extended further and would require more fixes than simply those mentioned above in relation to the initial Pressing Intensity metric.

\noindent
At some point these types of calculations will simply converge back to the more complex Pitch Control model by Spearman.
\subsection{Smart Pressing}
\begin{figure}[!ht]
    \centering
    \includegraphics[width=0.9\linewidth]{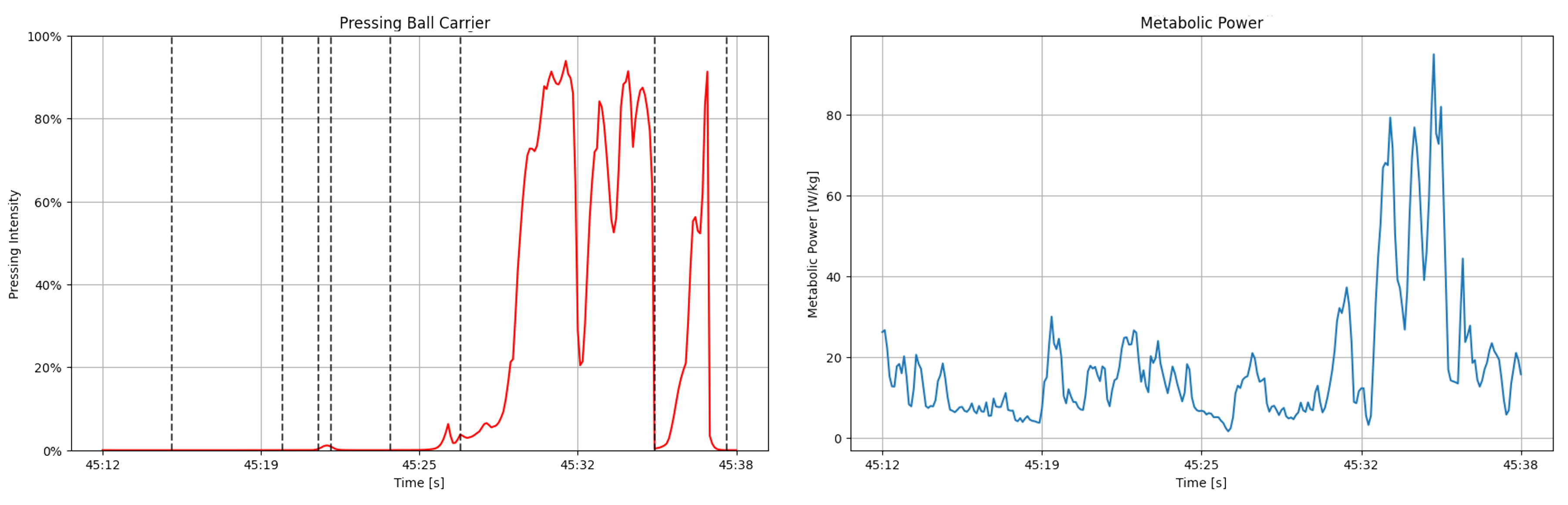}
    \caption{\centering Pressure on the ball carrier by some player (left) and their expended energy during the same time period (right)}
    \label{fig:mp}
\end{figure}

Because \textit{Pressing Intensity} is measured at the individual tracking frame level we could try to measure smart pressing, or efficient pressing, by considering the amount of energy expended during an out of possession sequence in the form of Metabolic Power ~\cite{Raabe22}. Figure~\ref{fig:mp} (on the left) shows a players’ Pressing Intensity on the ball carrier, where dashed lines indicating a change in ball carrier. On the right we show their metabolic power expended during this same time frame.

\begin{appendices}
\end{appendices}


\begin{thebibliography}{12}
\providecommand{\natexlab}[1]{#1}
\providecommand{\url}[1]{\texttt{#1}}
\expandafter\ifx\csname urlstyle\endcsname\relax
  \providecommand{\doi}[1]{doi: #1}\else
  \providecommand{\doi}{doi: \begingroup \urlstyle{rm}\Url}\fi

\bibitem[Andrienko et~al.(2017)Andrienko, Andrienko, Budziak, Dykes, Fuchs, Von~Landesberger, and Weber]{andrienko17}
Gennady Andrienko, Natalia Andrienko, Guido Budziak, Jason Dykes, Georg Fuchs, Tatiana Von~Landesberger, and Hendrik Weber.
\newblock Visual analysis of pressure in football.
\newblock \emph{Data Mining and Knowledge Discovery}, 31:\penalty0 1793--1839, 2017.

\bibitem[Bauer and Anzer(2021)]{bauer21}
Pascal Bauer and Gabriel Anzer.
\newblock Data-driven detection of counterpressing in professional football: a supervised machine learning task based on synchronized positional and event data with expert-based feature extraction.
\newblock \emph{Data Mining and Knowledge Discovery}, 35\penalty0 (5):\penalty0 2009--2049, 2021.

\bibitem{bekkers2024graph}
Bekkers, J., \& Sahasrabudhe, A. (2024). A Graph Neural Network deep-dive into successful counterattacks. \textit{arXiv preprint arXiv:2411.17450}.

\bibitem[Fernandez and Bornn(2018)]{fernandez18}
Javier Fernandez and Luke Bornn.
\newblock Wide open spaces: A statistical technique for measuring space creation in professional soccer.
\newblock In \emph{Sloan sports analytics conference}, volume 2018, 2018.

\bibitem[Merckx et~al.(2021)Merckx, Robberechts, Euvrard, and Davis]{merckx21}
Simon Merckx, Pieter Robberechts, Yannick Euvrard, and Jesse Davis.
\newblock Measuring the effectiveness of pressing in soccer.
\newblock In \emph{Workshop on Machine Learning and Data Mining for Sports Analytics}, 2021.

\bibitem[Morgan(2018)]{morgan18}
Will Morgan.
\newblock How statsbomb data helps measure counter-pressing, May 2018.
\newblock URL \url{https://statsbomb.com/articles/soccer/how-statsbomb-data-helps-measure-counter-pressing/}.

\bibitem[Osmanbasic(2017)]{basic17}
Adin Osmanbasic.
\newblock Pressing, counterpressing and counterattacking, 2017.
\newblock URL \url{https://spielverlagerung.com/2017/03/05/pressing-counterpressing-and-counterattacking/}.

\bibitem[Pleuler(2020)]{pleuler20}
Devin Pleuler.
\newblock Analytics handbook, 2020.
\newblock URL \url{https://github.com/devinpleuler/analytics-handbook}.
\newblock GitHub repository.

\bibitem[Raabe et~al.(2022)Raabe, Biermann, Bassek, Wohlan, Komitova, Rein, Groot, and Memmert]{Raabe22}
Dominik Raabe, Henrik Biermann, Manuel Bassek, Martin Wohlan, Rumena Komitova, Robert Rein, Tobias~Kuppens Groot, and Daniel Memmert.
\newblock floodlight - a high-level, data-driven sports analytics framework.
\newblock \emph{Journal of Open Source Software}, 7\penalty0 (76):\penalty0 4588, 2022.
\newblock \doi{10.21105/joss.04588}.
\newblock URL \url{https://doi.org/10.21105/joss.04588}.

\bibitem[Robberechts(2019)]{robberechts19}
Pieter Robberechts.
\newblock Valuing the art of pressing.
\newblock In \emph{StatsBomb Innovation in Football Conference}, volume~11, 2019.

\bibitem[Shaw(2020)]{shaw20}
Laurie Shaw.
\newblock Laurieontracking, 2020.
\newblock URL \url{https://github.com/Friends-of-Tracking-Data-FoTD/LaurieOnTracking}.
\newblock GitHub repository.

\bibitem[Spearman et~al.(2017)Spearman, Basye, Dick, Hotovy, and Pop]{spearman17}
William Spearman, Austin Basye, Greg Dick, Ryan Hotovy, and Paul Pop.
\newblock Physics-based modeling of pass probabilities in soccer.
\newblock In \emph{Proceeding of the 11th MIT Sloan Sports Analytics Conference}, volume~1, 2017.

\bibitem[Trainor(2014)]{trainor14}
Colin Trainor.
\newblock Defensive metrics: Measuring the intensity of a high press, July 2014.
\newblock URL \url{https://statsbomb.com/articles/soccer/defensive-metrics-measuring-the-intensity-of-a-high-press/}.

\bibitem{unravelsports}
Bekkers, J. (2024). unravelsports [GitHub]. Retrieved September 30, 2024, from \url{https://github.com/UnravelSports/unravelsports}


\end{thebibliography}
\end{document}